\documentclass{article}
\usepackage[T1]{fontenc} 
\usepackage[utf8]{inputenc} 
\usepackage{ismir,amsmath,cite,url}
\usepackage{graphicx}
\usepackage{color}
\def\lbd{}	

\usepackage[bookmarks=false,hidelinks]{hyperref}

\newcommand{\para}[1]{\vspace{1.5mm}\noindent\textbf{#1}.}

\title{Visual Guitar Tab Comparison}

\multauthor
{Frank Heyen \hspace{1cm} Alejandro Gabino Diaz Mendoza} {\bfseries{Quynh Quang Ngo \hspace{1cm} Michael Sedlmair }\\
VISUS, University of Stuttgart, Germany\\
{\tt\small \{frank.heyen, quynh.ngo, michael.sedlmair\}@visus.uni-stuttgart.de}
}

\def\authorname{F. Heyen, A. G. Diaz Mendoza, Q. Q. Ngo, and M. Sedlmair}

\begin{document}

\maketitle
\begin{abstract}
We designed a visual interface for comparing different guitar tablature (tab) versions of the same piece.
By automatically aligning the bars of these versions and visually encoding different metrics, our interface helps determine similarity, difficulty, and correctness. 
During our design, we collected and integrated feedback from musicians and finally conducted a qualitative evaluation with five guitarists.
Results confirm that our interface effectively supports comparison and helps musicians choose a version appropriate for their personal skills and tastes.
Our source code and online demo are available at \href{https://github.com/visvar/visual-guitar-tab-comparison}{github.com/visvar/visual-guitar-tab-comparison}.
\end{abstract}
\section{Introduction}\label{sec:introduction}

Many guitarists learn to play new pieces from tablature files available on the internet.
Such files are abundant on websites such as \href{https://www.ultimate-guitar.com/}{ultimate-guitar.com}, often with multiple variants or even edit histories.
While these different versions sometimes have user ratings, a musician looking for a tab can not always assume that the best-rated or most popular one is the most appropriate for them:
Maybe the most correct version exceeds their skills, 
maybe they prefer the sound of another one, 
or maybe the popularity and rating could be misleading, for example, when the most correct version has been added later and therefore gotten less attention and fewer ratings.
To support guitarists in this choice, we aim to provide them with a better understanding of the specific characteristics of each version, as well as explicitly revealing differences~\cite{gleicher2011visual} -- by visualizing them.
%
Music visualization is a diverse field~\cite{khulusi2020survey, lima2021musvistechs, miller2022corpusvis} and visually augmented sheet music helps analyze rhythm and harmonic patterns~\cite{ miller2022augmenting} or identifying mistakes in composition~\cite{deprisco2017understanding} and instrument practice~\cite{hori2019piano:hmm}.
Our work uses a similarity-based coloring~\cite{heyen2023visual} designed to reveal repeating structures in sheet music.
However, to the best of our knowledge, there is no visualization for comparing alternative versions.

\section{Design}

Tabs usually contain multiple instruments.
To make comparison feasible, we restrict our visualizations to only a single instrument the user chooses for each version.
Following the principle of focus and context, we designed two main views.
Our tab based on alphaTab (\href{https://www.alphatab.net/}{alphatab.net}) displays all bars of an instrument in a single row, allowing us to place all versions from top to bottom, such that the user can scroll through all at the same time.
An additional overview uses the same layout but reduces each bar to a small rectangle to fit the entire length of all versions onto the screen.
Serving as a minimap, this overview also lets users click on a bar to navigate the detailed tab view.
Since there might be missing or extra bars in some versions, we automatically align~\cite{needleman1970align} them by inserting empty bars.
Based on feedback from guitarists, we chose different metrics that we visualize using colors in both views:

\para{Note density}
With our density metric, we visualize how many notes each bar has. 
Differences between versions can hint at varying difficulty or correctness, for example, when chords in one version have more notes.

\begin{figure}[ht]
 \centerline{
 \includegraphics[width=0.9\columnwidth]{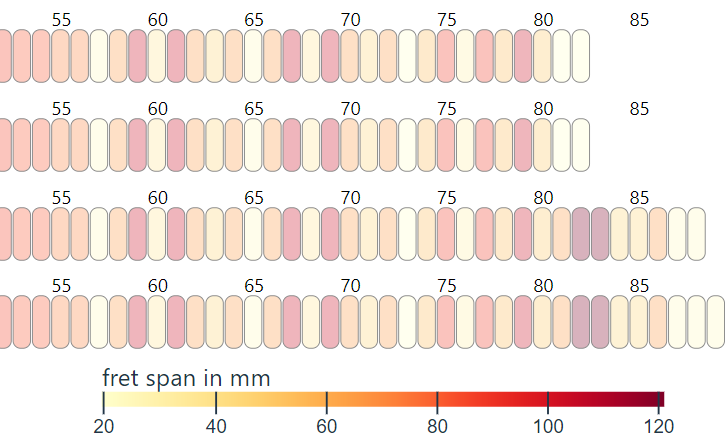}}
 \caption{The fret span in mm indicates how much the fretting hand has to move while playing each bar.}
 \label{fig:fretspan}
\end{figure}

\para{Fret span}
Another metric takes the lowest and highest fret played in each bar to determine how much the guitarist's fingers have to move while fretting notes.
We added an alternative, where we calculate the distance in millimeters instead of the number of frets (\autoref{fig:fretspan}).
This value is mapped to a color to facilitate comparison between but also within versions.

\vspace{10mm}

\begin{figure}[ht]
 \centerline{
 \includegraphics[width=\columnwidth]{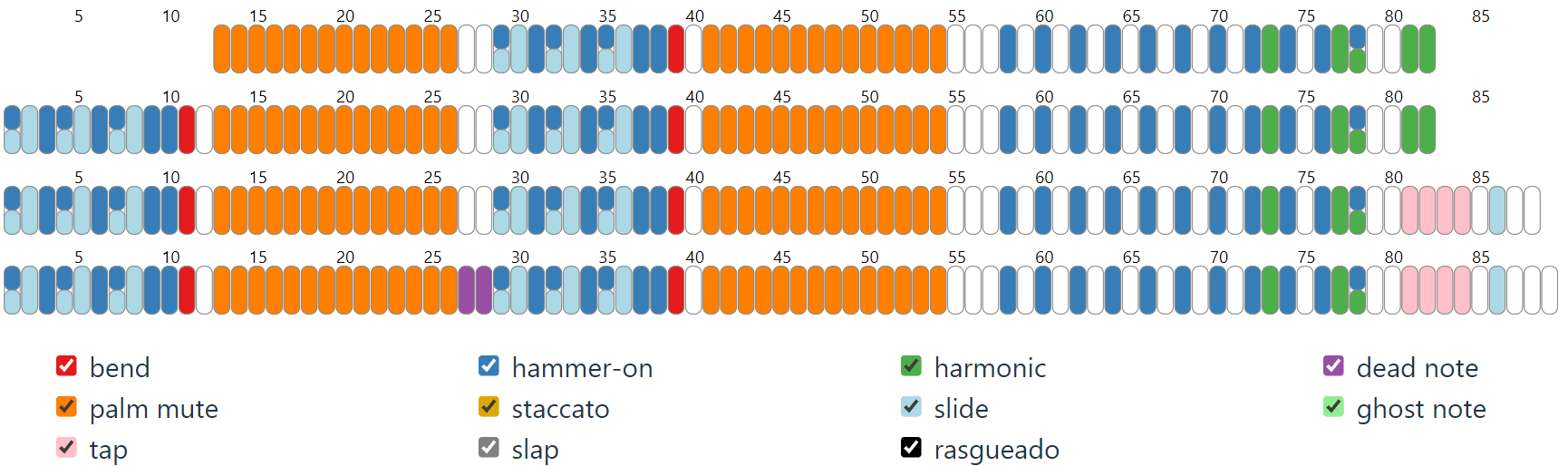}}
 \caption{Visualization of playing techniques used in each bar of each version.}
 \label{fig:techniques}
\end{figure}

\para{Playing techniques}
The annotation of different playing techniques, such as bends, palm muting, or harmonics, might vary between tab versions.
To help musicians spot versions with techniques they want to learn or avoid, we designed a visualization with color-coded stacked rectangles that encode the techniques used in each bar (\autoref{fig:techniques}).

\begin{figure}[ht]
 \centerline{
 \includegraphics[width=\columnwidth]{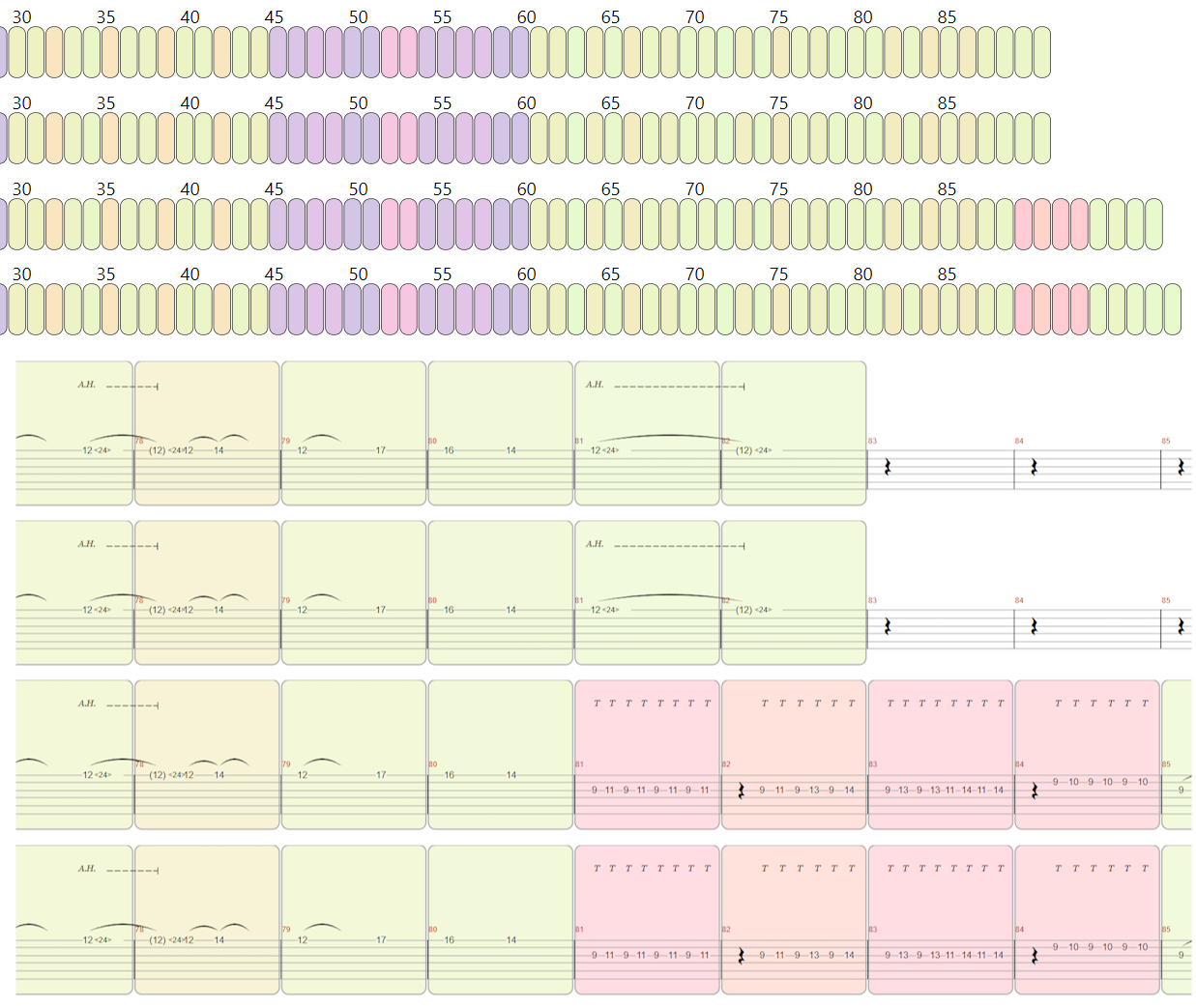}}
 \caption{Similar colors indicate similar bars.}
 \label{fig:similarity}
\end{figure}

\para{Bar similarity}
We further wanted to show the similarity between bars, within and between versions.
To indicate this similarity through colors, we apply an existing similarity-based color mapping~\cite{heyen2023visual} to the bars of all versions at once (\autoref{fig:similarity}).

\begin{figure}[ht]
 \centerline{
 \includegraphics[width=\columnwidth]{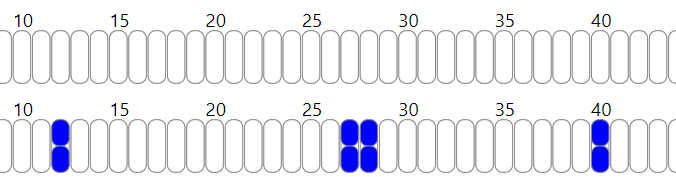}}
 \caption{Explicitly encoded changes (blue) between the top version and others.}
 \label{fig:directcomparison}
\end{figure}

\para{Explicit differences}
To reveal minor changes, for example, those that occur when someone uploads a slightly corrected version to a website, we explicitly encode differences (\autoref{fig:directcomparison}).
Such small edits could be barely visible in our other visualizations.

\section{Qualitative Evaluation}

We conducted a pair analytics study~\cite{arias2011pairanalytics} with five guitarists (P1--P5) who had no visualization background, where we spent one hour with each looking at tabs of their choice together.

\para{Note density}
To increase difficulty gradually, P1 would use a version with fewer notes first and, once learned, one with more notes later.
As P5 likes singing along, she prefers low-density versions where she does not have to play as much.

\para{Fret span}
P4, who plays bass guitar, found our metric for fret span in millimeters helpful, especially for beginners who cannot yet stretch their hands enough for the wider gaps between lower frets.

\para{Playing techniques}
Visualizing playing techniques in each bar of each version helped P2 determine which ones were more accurate, as he knew what the song should sound like.
P3 compared the use of techniques between versions to exclude outliers.
According to P5, seeing which techniques a tab contains could also increase the motivation to play it, as you know which ones you will practice or avoid.

\para{Bar similarity}
Having an overview of the structure could help him learn a new song faster mentioned P2, as he was able to detect verse, chorus, and bridge just by looking at the colors.
P3 would use similarity to spot and discard inaccurate versions, as they are more different from the rest.

\para{Explicit differences}
P3 suggested using both bar similarity and explicit differences together, as they complement each other: the former shows overall structure, and the latter highlights finer details.
When making his own arrangements, P4 sometimes lost track of changes -- changes he could easily spot with our interface.

\section{Conclusion}
Our proposed visual approach effectively supports comparison between versions of the same piece.
As our evaluation indicated, guitarists without prior visualization experience were able to make helpful findings and even created their own strategies for estimating quality and difficulty.
%
As a work in progress, our design also faces limitations:
The alignment does not work precisely for specific cases, for example, when the longest version misses bars that another one has -- such cases require multi-sequence alignment.
Participants missed audio playback to judge how far versions sound different.
Some techniques, such as sweep picking, might not be annotated in tab files and would require heuristics to detect.
Currently, there is no automatic recommendation or guidance, a feature that could personalized for taste or skill.
%
In the future, we want to address these limitations and refine our metrics to better handle chords and different playing techniques.
An extension to large collections of tabs could further support finding suitable pieces or exercises based on taste and skill.


\section{Acknowledgments}
This work was funded by the Cyber Valley Research Fund and by the Deutsche Forschungsgemeinschaft (DFG, German Research Foundation) -- Project-ID 251654672 -- SFB TRR 161, project A08.

\bibliography{ISMIRtemplate}

%
%
%
%
%

\end{document}